# Correlation of the estimated arrival time of the relativistic solar ions at 1 AU and start of ground level enhancement (GLE)

H.Martirosyan, A. Chilingarian.
*(a) Alikhanyan Physics Institute, Yerevan, Armenia, Alikhanyan Brothers 2, Yerevan 375036, Armenia*
*Presenter: A.Chilingarian (chili@crdlx5.yerphi.am), arm-martirosyan-H-abs1-sh35-poster*

We are investigated possible correlations between the calculated arrival times of the first relativistic ions at Earth and GLE start times registered by surface monitors. The analysis is based on the arrival times and energies of the first solar ions, registered by the Solar Isotope Spectrometer (SIS) on board of the ACE satellite, and protons, registered by GOES satellites.
We consider both cases when the interplanetary propagation of the first high energy ions is essentially scatter-free and the diffusion of high energy ions during propagation in the interplanetary magnetic field. We extrapolate the time-velocity and time-rigidity relationships to calculate the expected arrival times of the relativistic ions that are energetic enough to enter the atmosphere at the Aragats geographical location and produce secondary fluxes that reach the monitors.

## 1. Introduction

The spectrometric information from first satellites prove that Solar Energetic Particles (SEP) consists of protons and heavier ions. [1,2,3]. The energy spectra of solar ions are measured precisely by the facilities of ACE space station and GOES satellites till energies up to 1 GeV/nuclon. During the strongest of satellite era SEP event from 29 October 1989, the iron spectra at high energies was harder than proton spectra and iron ions contribution to the detected worldwide Ground Level Enhancement (GLE) was significant [4,5]. Using the method proposed in [6], we estimate expected arrival times of solar ions by the energy spectra measured by ACE/SIS [7] for the GLE from April 15 2001 [8,9]. Comparisons of obtained arrival times with monitors of Aragats Space Environmental Center (ASEC) [10] prove that heavy ions contribute to the GLE. In the report we present results on the GLE from 20 January 2005.

## 2. The proton fluxes at 20 January 2005

At 20 January, as for several other SEP events (April15, 2001, December 26, 2001) all energy channels of GOES satellite register enhancement of proton flux simultaneously. The SEP start was registered at 6:50 UT, 3 minutes before maximum of X-ray flare. 2005г [11]. Therefore, we can’t apply the regression method [6] to estimate the arrival times of first protons. On the other hand, the time history of changing spectra index and knee position, Figure 1, give valuable information on SEP origin.

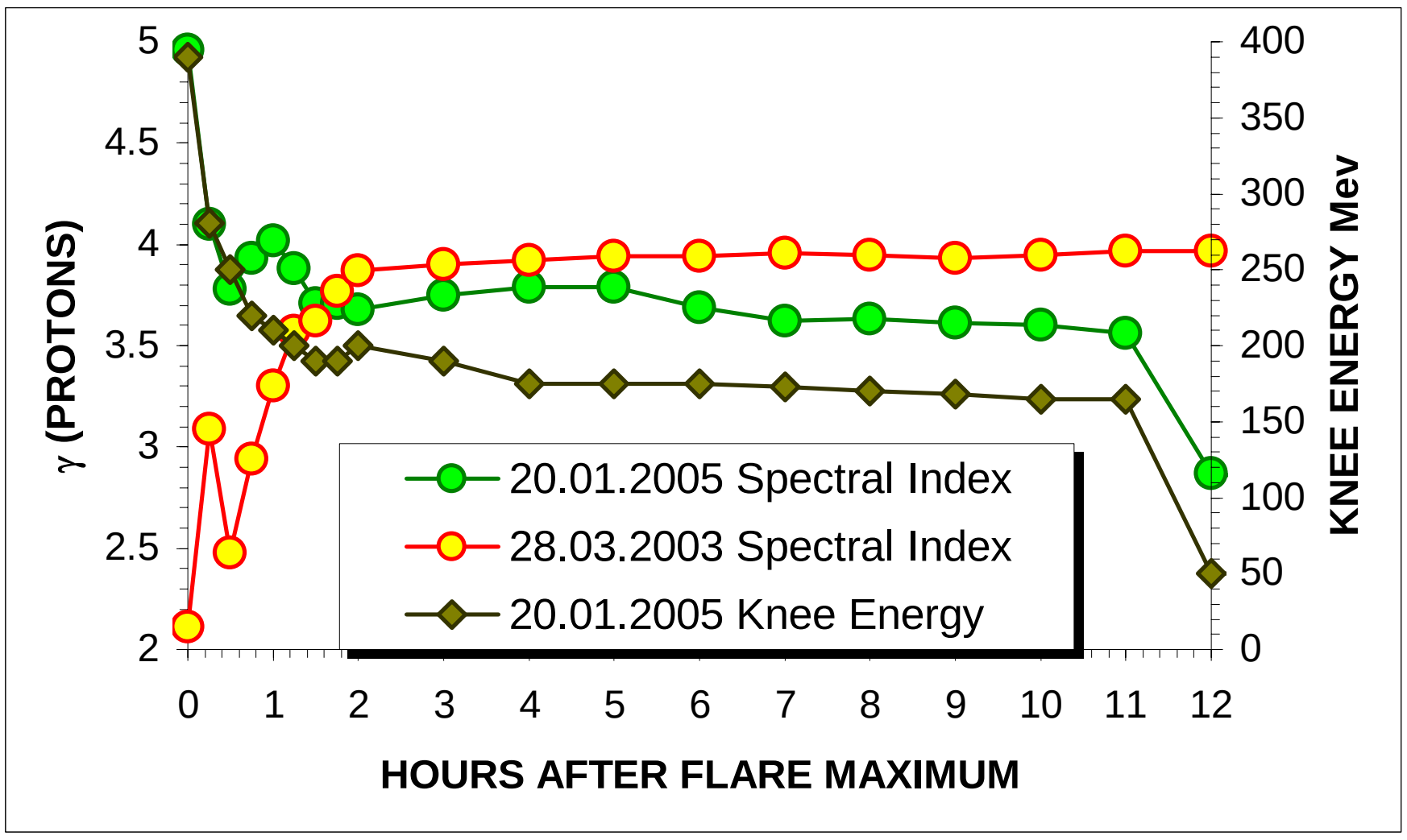

**Figure 1 The time history of the spectral index and knee position of proton spectrum 20 January 2005. For comparison the time history of spectral index for the 28 October 2003 is also posted**

During the first hour the proton spectrum parameters are changing dramatically. Such fast changes can't be connected with particle transport in the interplanetary space. The drastic changes of the spectra can be connected with some yet unknown particle acceleration mechanism preceding the CME launch (6:54 UT); also the time-span length is well-matched with flare duration. Therefore, we can assume that the high energy proton spectrum in the first hour is formed directly in the coronal flare site. Next 10 hours acceleration produces spectra with very stable parameters, most likely formed at CME driven shock site. It is interesting to note, that behavior of proton spectra of 28 October was vice-versa. Therefore the ion spectra can be used for classification of SEP events.

## 3. Ion spectra

In contrast to protons, first SEP ions arrive to Earth with relative delays inversely proportional to their energies, giving possibility to estimate the expected arrival times of GLE particles (>1GeV/nuclon) The results are posted in Figure 2; we use spectra measured by ACE/SIS [7].

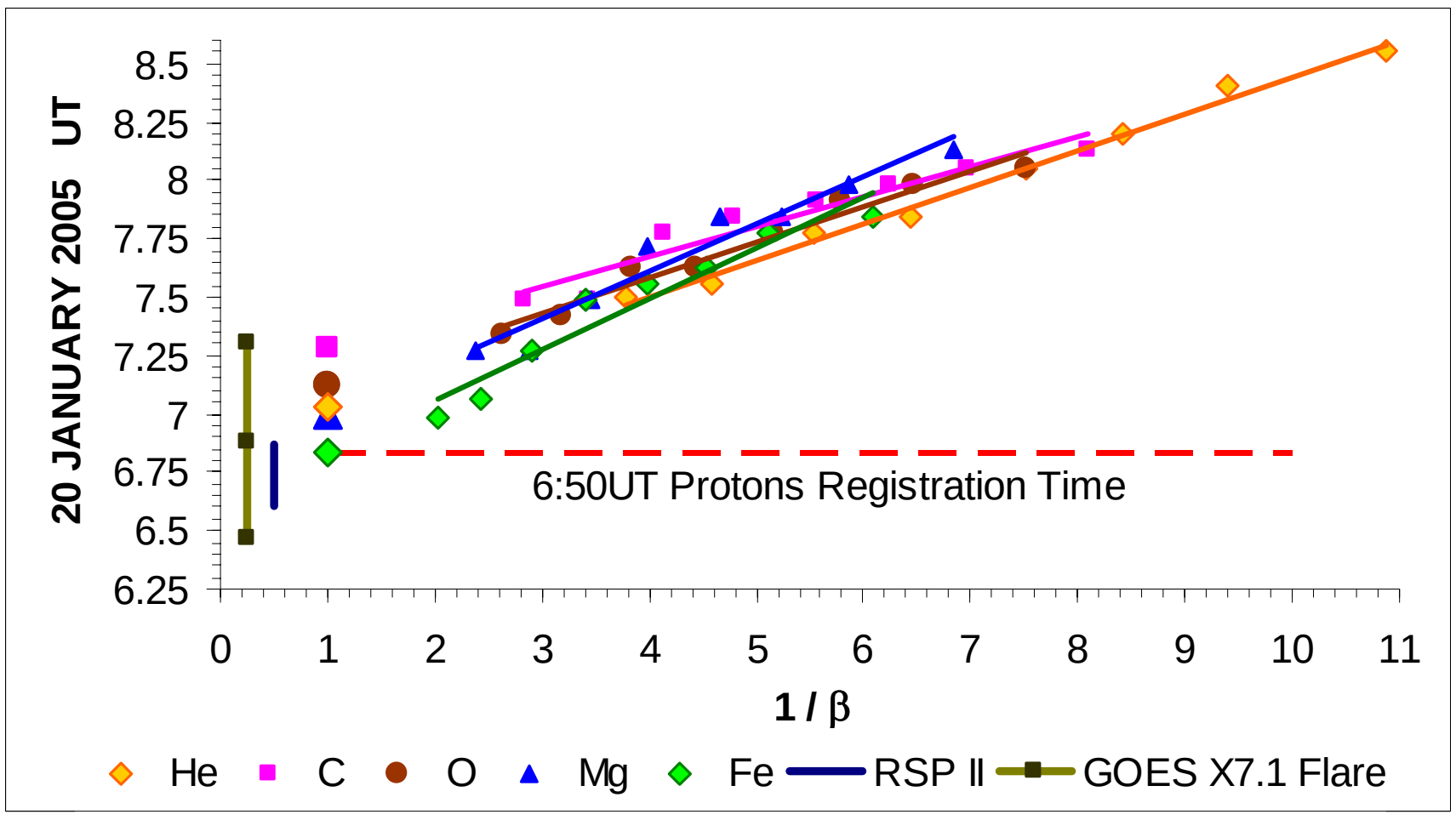

**Figure 2 Expected arrival times of GLE ions; 20 January 2005.**

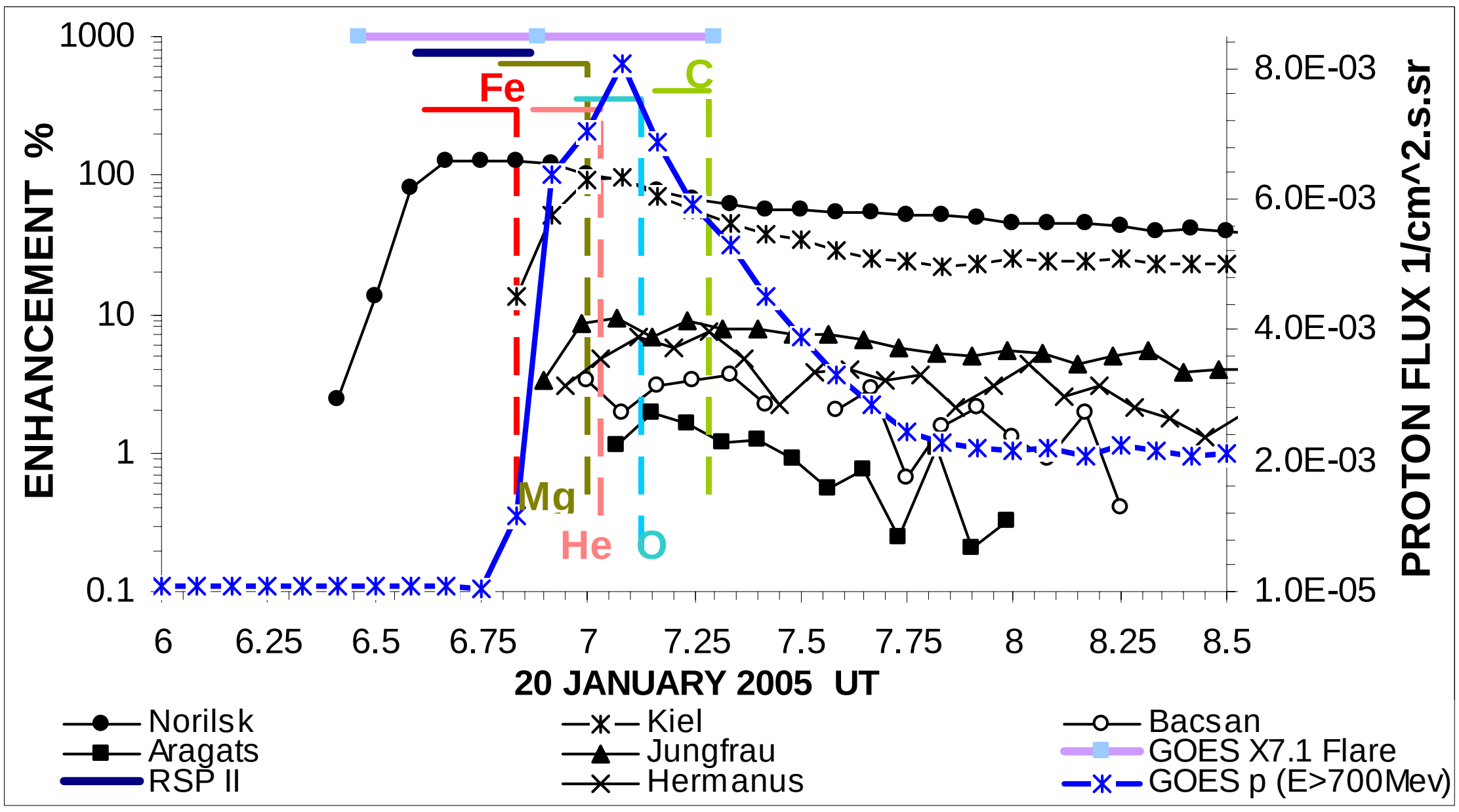

**Figure 3 Expected arrival times of GLE ions, superimposed on the GLE time profiles, measured by the Norilsk (0.63 GV), Jumgfraujoch (4.48 GV), Hermanus (4.9 GV), Baksan (5.6 GV) and Aragats (7.6 GV) Neutron Monitors; the error bars did not exceed 3-5 minutes**

Ion arrival expected times are denoted by vertical dashed lines. Horizontal lines connected to vertical ones on the top enumerate the ion flight duration and beginning of horizontal lines – the ion injection times. The ion injection times well fit the flare time. Expected arrival time of the first Fe ions coincide with proton arrival and start of GLE for the NM with cut-off rigidity less than 5 GV. The expected arrival times of first He, O and Mg ions coincide with GLE start for NM with cut-off rigidity greater than 5 GV. Inference from Figure 3 should be checked with SOHO/LASCO data on CME trajectory not available on the moment of report writing.

## 4. Discussion

SEP from 20 January is fastest in 23-rd solar cycle relative to X-ray flare and most mysterious from the point of view of the acceleration time and site. First protons of all energies arrive to 1 AU after 22 minutes from flare start (6:50 – 6:28 = 22 min.). Taking into account flare co-ordinates N12W58, and solar wind mean speed 840 km/sec we can estimate the length of particle path to reach Earth to be ~1,04AU.

Collisionless transport of protons with energies less than 80 MeV will take more than 22 minutes; therefore the low energy protons should be injected ***before start of X-ray flare*** (10 MeV protons should be injected half-of-hour before start of flare). This inference is confirmed by the Norilsk NM posted in Figure 3: the start of GLE coincides with start of X-ray flare (6:28 UT). The same behavior was observed for the SEP events from 15 April 2001 and 8 November 2000. These observations contradict common theories of solar particle acceleration [12].

The GLE start observed by low latitude NM ($R_c$ > 5 GV), is 10 later than proton arrival and coincides with expected arrival times of He, O, and Mg ions. Therefore we can bind the maximal energy of unknown solar accelerator by 10-15 GeV.

## Acknowledgement

The authors wish to thank ACE and GOES experiments groups as well as groups of Norilsk, Kiel, Jungfrajoch., Hermanus, Baksan\ NM for posting the data on the Internet.
The data collected by the ASEC detectors are the property of the ASEC collaboration. The authors thank CRD staff for the fruitful collaboration. Work was supported by the Armenian government grants and by ISTC grants A1058.